\documentclass[11pt]{article}

\usepackage[utf8]{inputenc}
\usepackage[T1]{fontenc}
\usepackage{amsmath,amssymb,amsthm}
\usepackage{graphicx}
\usepackage{booktabs}
\usepackage{array}
\usepackage{multirow}
\usepackage{makecell}
\usepackage{enumitem}
\usepackage[colorlinks=true,linkcolor=blue,citecolor=red,urlcolor=blue]{hyperref}
\usepackage{geometry}
\usepackage{natbib}
\usepackage{setspace}
\usepackage{caption}
\usepackage{float}
\usepackage{subcaption}

\allowdisplaybreaks

\newtheorem{theorem}{Theorem}[section]

\newtheorem{proposition}[theorem]{Proposition}

\theoremstyle{definition}
\newtheorem{definition}[theorem]{Definition}
\newtheorem{assumption}[theorem]{Assumption}
\theoremstyle{remark}

\newcommand{\Z}{\mathbb{Z}}
\newcommand{\lam}{\lambda}
\newcommand{\tht}{\theta}
\newcommand{\Nk}{N}
\newcommand{\Loss}{L}
\newcommand{\cw}{c_w}
\newcommand{\cN}{c_N}
\newcommand{\kap}{\kappa}
\newcommand{\qacc}{q}
\newcommand{\hacc}{h}
\newcommand{\muA}{\mu_A}
\newcommand{\muI}{\mu_I}
\newcommand{\PA}{P_A}
\newcommand{\PI}{P_I}
\newcommand{\Wq}{W_q}
\newcommand{\Tsys}{T}
\newcommand{\rhoN}{\rho}

\title{\textbf{Liability Sharing and Staffing in AI-Assisted Online Medical Consultation}}

\author{
    Yang Xiao\thanks{Corresponding author. Email: \href{mailto:yang.xiao@saitama-u.ac.jp}{yang.xiao@saitama-u.ac.jp}}\\
    \small School of Business, Nantong University, Nantong, China\\
    \small Faculty of Economics, Saitama University, Saitama, Japan
}

\date{\today}

\begin{document}
\maketitle

\begin{abstract}
\noindent Liability sharing and staffing jointly determine service quality in AI-assisted online medical consultation, yet their interaction is rarely examined in an integrated framework linking contracts to congestion via physician responses. This paper develops a Stackelberg queueing model where the platform selects a liability share and a staffing level while physicians choose between AI-assisted and independent diagnostic modes. Physician mode choice exhibits a threshold structure, with the critical liability share decreasing in loss severity and increasing in the effort cost of independent diagnosis. Optimal platform policy sets liability below this threshold to trade off risk transfer against compliance costs, revealing that liability sharing and staffing function as substitute safety mechanisms. Higher congestion or staffing costs tilt optimal policy toward AI-assisted operation, whereas elevated loss severity shifts the preferred regime toward independent diagnosis. The welfare gap between platform and social optima widens with loss severity, suggesting greater scope for incentive alignment in high-stakes settings. By endogenizing physician mode choice within a congested service system, this study clarifies how liability design propagates through queueing dynamics, offering guidance for calibrating contracts and capacity in AI-assisted medical consultation.

\medskip
\noindent\textbf{Keywords:} AI-assisted medical consultation; liability sharing; physician staffing; Stackelberg game; queueing.

\end{abstract}

\newpage


\section{Introduction}

\subsection{Background and Practical Motivation}

The integration of generative artificial intelligence into online medical consultation workflows represents one of the most significant transformations in healthcare service delivery. Contemporary platforms increasingly deploy AI systems for pre-consultation triage, automated symptom collection, medical record drafting, and real-time physician decision support \citep{Topol2019, Rajkomar2019}. These technological deployments are frequently justified as capacity expansion mechanisms: when patient arrival volumes surge, platforms can sustain acceptable response times by delegating routine cases to AI processing or by enabling physicians to rapidly confirm algorithmically generated recommendations.

However, recent developments in practice reveal substantial operational frictions that complicate this optimistic narrative. Platform incentive structures may inadvertently create a reverse diagnostic process wherein commercial purchasing intent precedes rather than follows proper medical verification, fundamentally compromising clinical integrity. Significant transparency deficits persist as patients frequently cannot distinguish whether they are interacting with a human physician or an AI-mediated workflow, materially affecting trust formation and complaint incidence rates. Most critically, liability allocation remains deeply contested: the platform derives operational benefits from AI-assisted workflows that reduce average service time per case, while individual physicians bear substantial reputational and legal exposure when diagnostic errors occur.

This operational setting exhibits precisely the structural characteristics amenable to rigorous operations research analysis. Queueing theory establishes that system congestion increases sharply and nonlinearly as utilization approaches saturation. Contract theory demonstrates that liability allocation systematically influences agent behavior. When these two fundamental forces interact within a unified framework, ostensibly minor contractual modifications can generate substantial operational consequences amplified through queueing nonlinearity. This interaction effect motivates our integrated modeling approach.

\subsection{Problem Statement}

We investigate an online medical consultation platform that receives patient consultation requests over continuous time according to a stochastic arrival process. The platform maintains a pool of physicians to process these requests, and for each patient encounter, physicians can operate in one of two distinct diagnostic modes. In Mode A (AI-Assisted), the physician adopts AI-generated diagnostic recommendations and provides rapid confirmation, achieving faster service but accepting elevated error risk. In Mode I (Independent), the physician conducts comprehensive independent diagnosis without algorithmic assistance, requiring longer service time but achieving superior diagnostic accuracy.

The platform controls two primary decision levers: the liability share allocated to physicians when diagnostic errors occur, and the aggregate staffing level. Physicians respond strategically by selecting the diagnostic mode that maximizes their expected utility given the prevailing liability regime. The service system operates under congestion, creating interdependence between staffing decisions and mode choices through their joint effect on realized service rates. The platform seeks to minimize long-run expected total cost encompassing risk-related losses, congestion-induced patient experience degradation, staffing expenditures, and regulatory compliance costs.

\subsection{Research Questions}

This paper addresses two interconnected research questions. First, how does the liability sharing mechanism induce physicians to choose between rapid AI-assisted diagnosis and thorough independent diagnosis, and how should the platform jointly optimize liability sharing and staffing decisions to balance risk exposure and service congestion? Second, how do regulatory constraints on liability allocation shift optimal policy configurations and regime boundaries, and what is the magnitude of the welfare gap between platform-optimal policy and a social benchmark?

\subsection{Summary of Contributions}

This paper makes four substantive contributions to the service operations and healthcare management literatures. We develop an integrated analytical framework combining Stackelberg liability design with M/M/N queueing staffing, thereby endogenizing physician behavioral responses within a congested service system and addressing a fundamental gap between incentive-focused and capacity-focused research traditions. Our model employs discrete diagnostic mode selection rather than continuous effort choice, accurately reflecting the operational reality that platforms typically promote distinct workflow configurations rather than continuous effort gradations. We establish formal regime comparison methodology by defining scenario comparisons as constrained optimization problems with precisely specified feasible sets, enabling rigorous regime maps and interpretable boundary characterizations. Finally, we derive provable structural properties including threshold characterizations, convexity properties, and existence guarantees, while maintaining computational transparency through explicit enumeration procedures.

\subsection{Paper Organization}

The remainder of this paper proceeds as follows. Section~\ref{sec:literature} positions our work within three relevant research streams. Section~\ref{sec:model} presents the complete model specification. Section~\ref{sec:physician} analyzes physician mode choice. Section~\ref{sec:platform} develops platform optimization. Section~\ref{sec:scenarios} defines policy scenarios. Section~\ref{sec:numerical} presents numerical results and analysis. Section~\ref{sec:implications} discusses managerial implications. Section~\ref{sec:conclusion} concludes with discussion of limitations and future directions.

\section{Literature Review}
\label{sec:literature}

We position this study at the intersection of three established research streams: queueing-based staffing in service operations, human-AI collaboration and algorithmic reliance, and liability allocation in contracting relationships.

\subsection{Queueing Theory and Staffing in Service Operations}

Queueing theory and optimal staffing constitute foundational topics in service operations management, with particularly extensive development in call center operations and healthcare delivery systems \citep{GrossHarris, KooleMandelbaum, GansKooleMandelbaum}. A central insight from this literature is that expected waiting time exhibits pronounced nonlinearity near full capacity utilization, establishing staffing as a primary lever for service quality management. The seminal work of \citet{HalfinWhitt} established the square-root staffing principle for large-scale service systems, demonstrating that staffing levels should scale with the square root of offered load to balance quality and efficiency. \citet{BorstMandelbaumReiman} extended this framework to call center dimensioning, while \citet{GarnettMandelbaumReiman} incorporated customer impatience and abandonment. Healthcare operations research has applied queueing insights to appointment scheduling \citep{CayirliVeral, GuptaDenton}, operating room management \citep{CardoenDemeulemeesterBelien}, and patient flow optimization \citep{HallPatientFlow}. A significant limitation of the existing literature is the treatment of service rates and service quality as exogenous parameters. In AI-assisted medical consultation, the effective service rate depends endogenously on whether physicians adopt a rapid AI confirmation workflow versus conducting independent diagnosis. Our model addresses this limitation by making mode choice, and consequently service rate, responsive to contractual incentives.

\subsection{Human-AI Collaboration and Algorithmic Reliance}

A substantial behavioral literature examines how human decision makers interact with algorithmic recommendations. \citet{DietvorstSimmonsMassey} documented algorithm aversion whereby decision makers avoid algorithms after observing algorithmic errors, even when algorithms demonstrably outperform human judgment. Conversely, \citet{LoggMinsonMoore} identified conditions under which individuals exhibit algorithm appreciation, preferring algorithmic to human judgment. \citet{CasteloBosLehmann} demonstrated that algorithm aversion is task-dependent, with greater aversion for subjective tasks requiring human insight. In clinical decision support contexts, \citet{Topol2019} surveys the convergence of human and artificial intelligence in high-performance medicine, while \citet{Rajkomar2019} and \citet{Esteva2019} review machine learning applications across medical domains. A consistent finding is that the human clinician remains the final decision maker in most deployment contexts, implying that the operationally relevant question concerns not algorithm accuracy per se, but how system design influences human reliance decisions. The human-AI collaboration literature generally abstracts away from operational constraints including queueing dynamics and capacity decisions. Our model introduces the congestion channel: changes in algorithmic reliance alter effective service rates, which propagate through queueing dynamics to affect waiting times and demand-side experience costs.

\subsection{Liability Allocation and Contracting}

Economic analysis of liability rules and incentive design has a distinguished intellectual history. \citet{Shavell1980} established foundational results on strict liability versus negligence standards, demonstrating how liability rules shape precaution incentives and risk allocation. \citet{Shavell2004} provides comprehensive treatment of the economic analysis of law including tort liability. Contract theory examines risk sharing and incentive compatibility in principal-agent relationships, with \citet{Holmstrom1979} establishing the informativeness principle for optimal contracting and \citet{Sappington1991} surveying incentive design in regulatory contexts. \citet{LaffontMartimort} and \citet{BoltonDewatripont} provide comprehensive textbook treatments of modern contract theory. In platform-mediated services, liability allocation can be interpreted as an incentive lever affecting frontline worker behavior. However, the existing contracting literature rarely incorporates congestion externalities arising from capacity constraints. Our contribution lies in embedding liability design within an M/M/N service system where these interactions are explicitly modeled.

\subsection{Research Gap and Positioning}

The literature review reveals a clear gap: queueing staffing models capture congestion dynamics but treat service quality and service rates as exogenous, while liability contracting models capture incentive effects but ignore congestion externalities. This paper develops a framework that addresses both dimensions simultaneously, enabling analysis of how liability design and staffing decisions interact through physician behavioral responses and queueing dynamics.

\section{Model Description and Assumptions}
\label{sec:model}

\subsection{Players, Timing, and Information Structure}

We model a two-stage Stackelberg game between a platform (leader) and a pool of physicians (followers). In Stage 1, the platform chooses the physician liability share $\tht \in [0,1]$ and the number of physicians $\Nk \in \Z_+$. In Stage 2, observing the announced liability share $\tht$, each physician selects a diagnostic mode $m \in \{A, I\}$ for processing patient cases. We assume complete information: all model parameters are common knowledge, and the platform correctly anticipates physician responses when making Stage 1 decisions.

\begin{assumption}[Physician homogeneity]
\label{assum:homogeneous}
All physicians share identical service rates, accuracy levels, disutility parameters, and compensation. Consequently, all physicians select the same diagnostic mode in equilibrium.
\end{assumption}

\subsection{Patient Arrivals and Service System}

Patient consultation requests arrive according to a Poisson process with rate $\lam > 0$. The platform staffs $\Nk$ identical physicians operating in parallel, with each physician serving one patient at a time. Service times are exponentially distributed with mode-dependent rates.

\begin{assumption}[Service rate ordering]
\label{assum:service}
If physicians operate in mode $m \in \{A, I\}$, the service rate per physician is $\mu_m$ satisfying $\muA > \muI > 0$.
\end{assumption}

The ordering $\muA > \muI$ reflects the operational reality that AI-assisted confirmation is faster than comprehensive independent diagnosis. Under these assumptions, the service system constitutes an M/M/$\Nk$ queue with arrival rate $\lam$ and aggregate service rate $\Nk \mu_m$.

\subsection{Queueing Performance Measures}

Define the traffic intensity under mode $m$ as $\rhoN_m = \lam / (\Nk \mu_m)$. System stability requires $\rhoN_m < 1$, establishing a minimum staffing requirement $\Nk \geq \underline{\Nk}_m \equiv \lfloor \lam / \mu_m \rfloor + 1$. The Erlang C formula gives the probability that an arriving patient must wait:
\begin{equation}
    C(\Nk, \rho) = \frac{\displaystyle\frac{(\Nk\rho)^{\Nk}}{\Nk!} \cdot \frac{1}{1-\rho}}{\displaystyle\sum_{k=0}^{\Nk-1} \frac{(\Nk\rho)^k}{k!} + \frac{(\Nk\rho)^{\Nk}}{\Nk!} \cdot \frac{1}{1-\rho}}.
    \label{eq:erlangC}
\end{equation}

The expected waiting time in queue is $\Wq(\lam, \Nk, \mu_m) = C(\Nk, \rhoN_m) / (\Nk\mu_m - \lam)$, and the expected total system time is $\Tsys_m(\lam, \Nk) = \Wq(\lam, \Nk, \mu_m) + 1/\mu_m$.

\begin{figure}[H]
    \centering
    \includegraphics[width=0.75\textwidth]{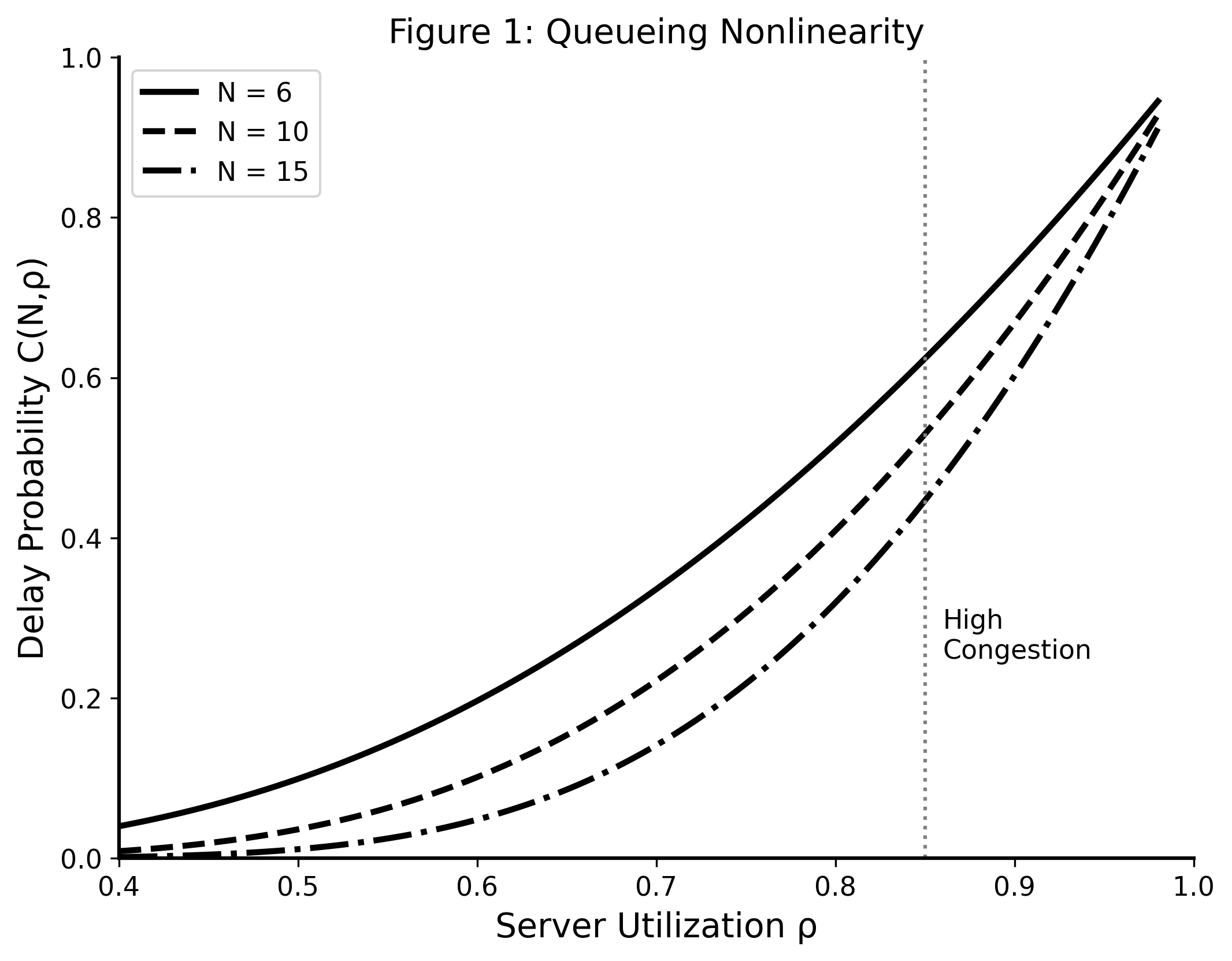}
    \caption{Queueing system nonlinearity illustrated through the Erlang C delay probability as a function of server utilization. The delay probability exhibits modest values for utilization below 0.8 but rises sharply as the system approaches saturation, creating an amplification mechanism whereby liability-induced mode switching generates disproportionate congestion effects.}
    \label{fig:queueing}
\end{figure}

Figure~\ref{fig:queueing} illustrates the fundamental nonlinearity of the Erlang C delay probability. The figure reveals that the delay probability remains relatively modest when server utilization is below 0.8, but rises precipitously as utilization approaches unity. This nonlinearity has profound implications for our model: when the system operates near capacity, small changes in effective service rate induced by physician mode switching can generate disproportionately large changes in patient waiting times. The three curves corresponding to different staffing levels ($N = 6, 10, 15$) demonstrate that while higher staffing reduces delay probability at any given utilization, the characteristic sharp rise near saturation persists regardless of scale. This queueing nonlinearity serves as an amplification mechanism that magnifies the operational consequences of liability-induced mode choice, establishing the foundation for our subsequent analysis of regime boundaries and optimal policy design.

\subsection{Diagnostic Accuracy and Risk Model}

Let $\Loss > 0$ denote the expected monetary loss when a diagnostic error occurs, encompassing direct medical costs, legal liability, reputational damage, and patient harm valuation.

\begin{assumption}[Accuracy ordering]
\label{assum:accuracy}
Define AI diagnostic accuracy as $\qacc$ and physician independent diagnostic accuracy as $\hacc$, satisfying $0 < \qacc < \hacc < 1$.
\end{assumption}

The mode-dependent error probabilities are $\PA = 1 - \qacc$ and $\PI = 1 - \hacc$, with $\PI < \PA$ confirming that independent diagnosis achieves superior accuracy. The parameter $\qacc$ represents diagnostic accuracy achieved when physicians rapidly confirm AI recommendations without independent verification, reflecting both AI system accuracy and physician reduced scrutiny under time pressure. The parameter $\hacc$ represents accuracy under careful independent diagnosis, reflecting physician expertise absent algorithmic assistance. The gap $\hacc - \qacc$ captures the accuracy cost of AI-assisted acceleration.

\subsection{Liability Allocation and Physician Utility}

The platform implements a linear liability sharing rule: when a diagnostic error occurs, physicians bear fraction $\tht \in [0,1]$ of the loss $\Loss$, while the platform bears fraction $(1 - \tht)$. Physicians receive fixed compensation $w > 0$ per unit time and incur mode-dependent operating disutility $k_m$ capturing cognitive effort, psychological burden, and professional discomfort.

\begin{assumption}[Disutility ordering]
\label{assum:disutility}
Mode-dependent disutility satisfies $0 < k_A < k_I$.
\end{assumption}

The ordering $k_A < k_I$ reflects that rapid AI confirmation requires less cognitive effort than comprehensive independent diagnosis. The physician's expected utility from operating in mode $m$ under liability share $\tht$ is:
\begin{equation}
    U_D(m; \tht) = w - k_m - \tht \Loss P_m.
    \label{eq:physician_utility}
\end{equation}

\subsection{Platform Cost Structure}

The platform's objective function aggregates four cost components expressed per unit time. Risk cost captures expected platform-borne loss: $C_{\text{risk}} = \lam (1-\tht) \Loss P_m$. Congestion cost reflects patient waiting and experience degradation: $C_{\text{congestion}} = \lam \cw \Tsys_m(\lam, \Nk)$, where $\cw > 0$ captures patient time valuation and satisfaction impact. Staffing cost represents physician employment expenditure: $C_{\text{staffing}} = \cN \Nk$. Compliance cost encompasses legal, administrative, and retention costs associated with liability shifting: $C_{\text{compliance}} = \kap \tht^2 \Nk$, where the quadratic form ensures interior optimal solutions and reflects increasing marginal difficulty of shifting liability.

The platform's total cost function is:
\begin{equation}
    TC(\tht, \Nk, m) = \lam(1-\tht)\Loss P_m + \lam\cw\Tsys_m(\lam,\Nk) + \cN\Nk + \kap \tht^2 \Nk.
    \label{eq:platform_cost}
\end{equation}

\subsection{Calibrated Parameter Values}

Table~\ref{tab:parameters} presents the calibrated parameter values used throughout our analysis. These parameters are selected to achieve meaningful regime structure with both AI-assisted and independent diagnosis regimes represented in the parameter space, and to ensure that the optimal liability share constitutes an interior solution rather than a boundary value.

\begin{table}[H]
    \caption{Calibrated parameter values and interpretation.}
    \label{tab:parameters}
    \centering
    \small
    \renewcommand{\arraystretch}{1.2}
    \begin{tabular}{@{}llll@{}}
        \toprule
        \textbf{Parameter} & \textbf{Value} & \textbf{Range} & \textbf{Interpretation} \\
        \midrule
        $\lam$ & 50/hour & [25, 90] & Patient arrival rate \\
        $\muA$ & 12/hour & Fixed & 5-minute AI-assisted service \\
        $\muI$ & 6/hour & Fixed & 10-minute independent service \\
        $\qacc$ & 0.90 & [0.80, 0.94] & AI diagnostic accuracy \\
        $\hacc$ & 0.95 & Fixed & Physician diagnostic accuracy \\
        $\Loss$ & 2,000 & [800, 5,000] & Expected loss per error \\
        $\cw$ & 150 & [50, 200] & Hourly waiting cost \\
        $\cN$ & 200 & [100, 350] & Hourly physician cost \\
        $\kap$ & 2,500 & [1,000, 5,000] & Compliance cost coefficient \\
        $k_A$ & 50 & Fixed & Mode A disutility \\
        $k_I$ & 110 & Fixed & Mode I disutility \\
        \bottomrule
    \end{tabular}
\end{table}

Under these calibrated parameters, the physician threshold is $\tht^D = (k_I - k_A) / (\Loss(\hacc - \qacc)) = 60 / (2000 \times 0.05) = 0.60$, and the optimal platform policy achieves $\tht^* = 0.40$ with $\Nk^* = 5$ physicians under AI-assisted mode (Regime A).

\section{Physician Mode Choice and Threshold Structure}
\label{sec:physician}

\subsection{Best Response Derivation}

Physicians choose mode A if and only if $U_D(A; \tht) \geq U_D(I; \tht)$. Substituting from \eqref{eq:physician_utility} and simplifying yields the threshold condition $k_I - k_A \geq \tht \Loss (\hacc - \qacc)$. Define the physician threshold:
\begin{equation}
    \tht^D = \frac{k_I - k_A}{\Loss(\hacc - \qacc)}.
    \label{eq:threshold}
\end{equation}

\begin{proposition}[Physician best response]
\label{prop:best_response}
The physician best response exhibits threshold structure: $m^*(\tht) = A$ if $\tht \leq \tht^D$, and $m^*(\tht) = I$ if $\tht > \tht^D$.
\end{proposition}

\begin{figure}[H]
    \centering
    \includegraphics[width=0.8\textwidth]{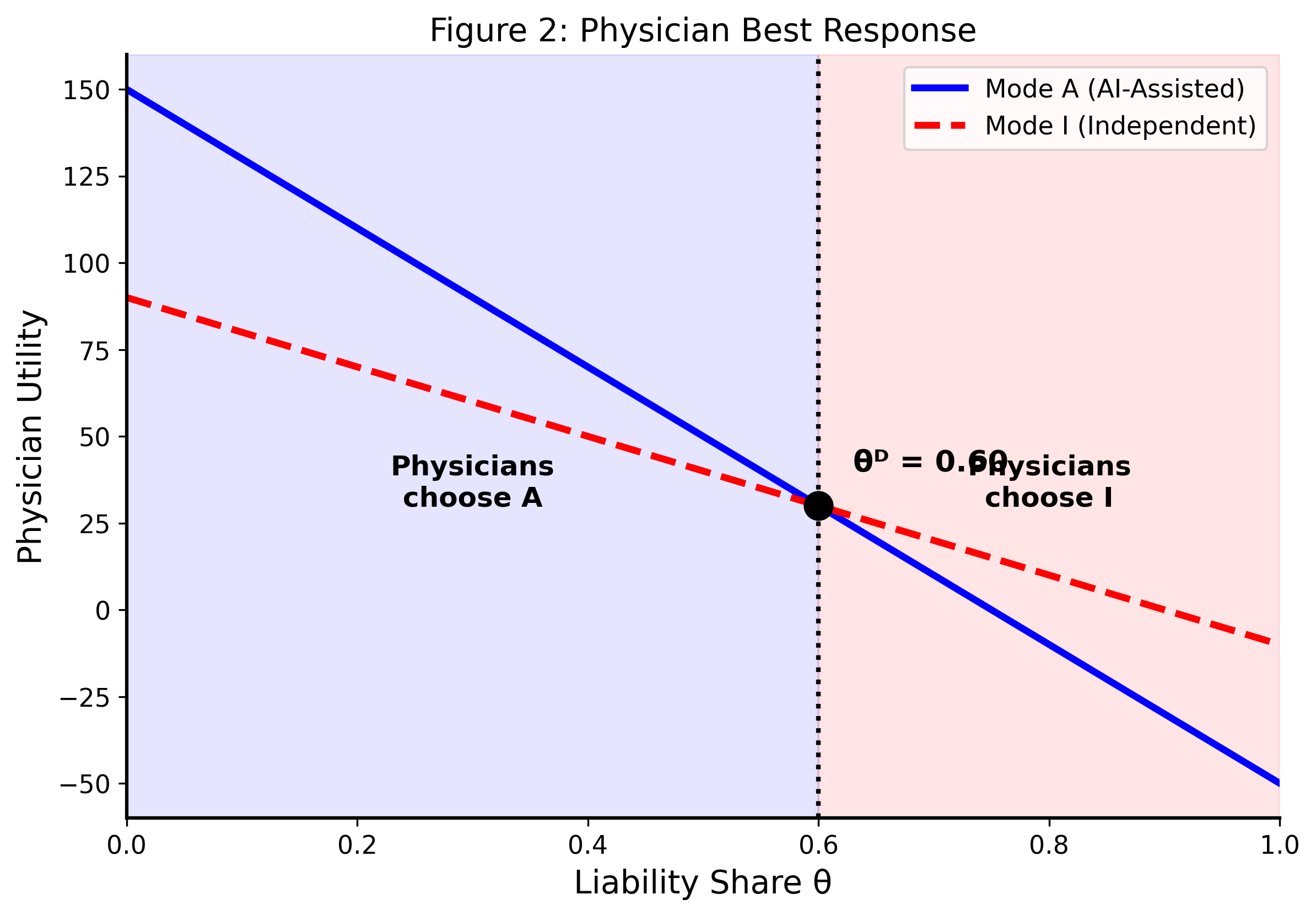}
    \caption{Physician best response and threshold structure. The intersection of Mode A and Mode I utility curves at $\tht^D = 0.60$ determines physician mode selection. Below the threshold, effort savings dominate liability costs, inducing AI-assisted mode; above the threshold, liability costs dominate, inducing independent diagnosis.}
    \label{fig:physician}
\end{figure}

Figure~\ref{fig:physician} provides a geometric interpretation of the physician threshold structure. The figure displays physician expected utility as a function of liability share for both diagnostic modes. The Mode A utility curve (solid blue) begins higher at $\tht = 0$ due to lower effort cost $k_A$, but declines more steeply because Mode A carries higher error probability $P_A = 1 - \qacc = 0.10$. The Mode I utility curve (dashed red) starts lower reflecting higher effort cost $k_I$, but its gentler slope reflects lower error probability $P_I = 1 - \hacc = 0.05$. The curves intersect precisely at the threshold $\tht^D = 0.60$, marked by the black dot. Below this threshold, physicians prefer Mode A because effort savings outweigh expected liability costs; above this threshold, the liability cost differential dominates, inducing the safer independent mode. The shaded regions visually demarcate the two behavioral regimes, illustrating how the platform can induce desired physician behavior through strategic liability allocation. This threshold structure is fundamental to our subsequent platform optimization analysis, as it partitions the platform's feasible set into distinct regimes with different operational characteristics.

\subsection{Comparative Statics of the Threshold}

\begin{proposition}[Threshold comparative statics]
\label{prop:comparative_statics}
The physician threshold $\tht^D$ satisfies: $\partial \tht^D / \partial \qacc > 0$, $\partial \tht^D / \partial \hacc < 0$, $\partial \tht^D / \partial \Loss < 0$, and $\partial \tht^D / \partial (k_I - k_A) > 0$.
\end{proposition}

\begin{figure}[H]
    \centering
    \includegraphics[width=\textwidth]{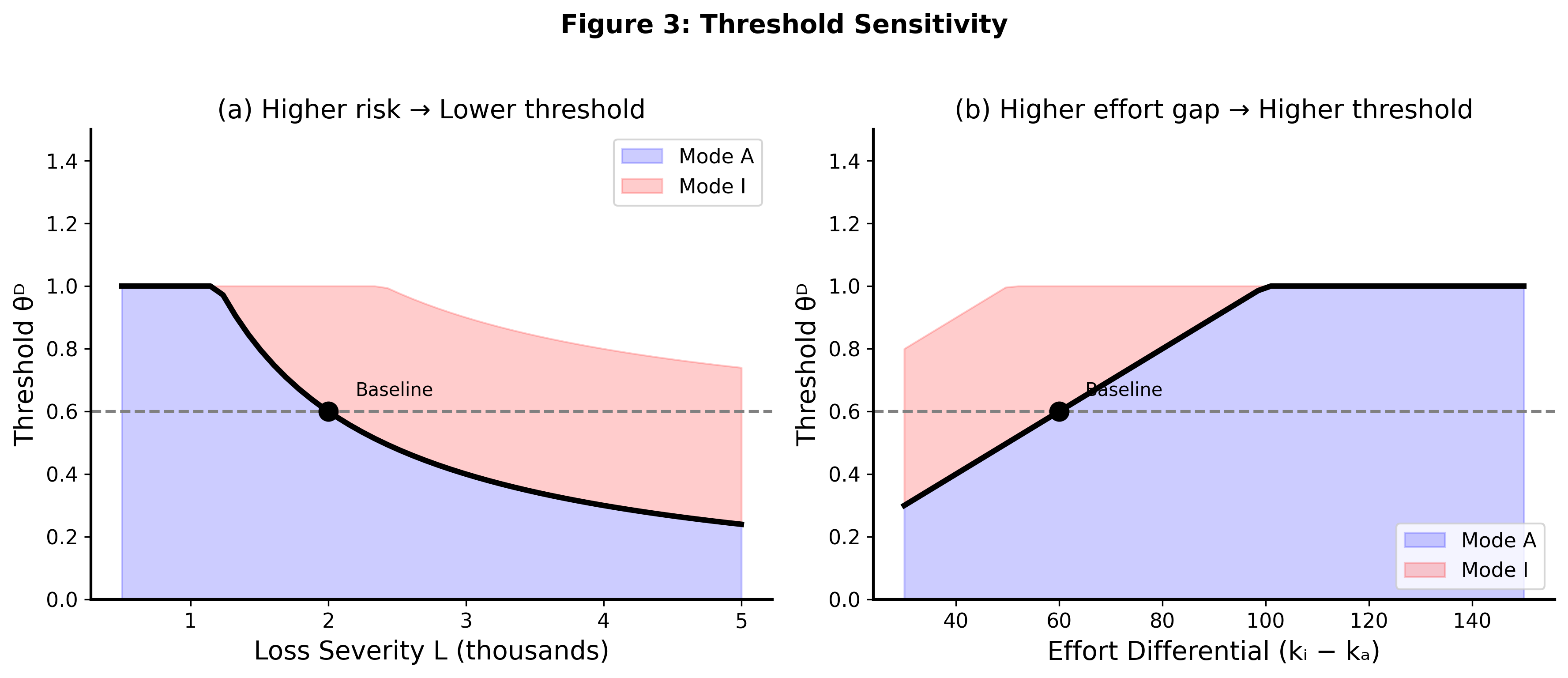}
    \caption{Sensitivity of physician threshold $\tht^D$ to model parameters. Panel (a): Higher loss severity $\Loss$ reduces the threshold, making physicians more responsive to liability incentives. Panel (b): Higher effort differential $k_I - k_A$ raises the threshold, requiring greater liability exposure to induce independent diagnosis.}
    \label{fig:threshold}
\end{figure}

Figure~\ref{fig:threshold} illustrates the sensitivity of the physician threshold to two key parameters. Panel (a) demonstrates that $\tht^D$ decreases hyperbolically in loss severity $\Loss$: when diagnostic errors carry severe consequences, even moderate liability shares induce physicians to adopt the safer independent mode. The baseline parameter $L = 2,000$ yields $\tht^D = 0.60$, but increasing loss severity to $L = 4,000$ would reduce the threshold to approximately 0.30, substantially expanding the region where independent diagnosis is optimal. Panel (b) shows that $\tht^D$ increases linearly in the effort differential $k_I - k_A$: when independent diagnosis imposes substantial additional cognitive burden relative to AI-assisted confirmation, physicians require higher liability exposure before switching to the more demanding mode. These comparative statics have direct policy implications. Regulators seeking to encourage safer diagnostic practices can either increase physician liability exposure (shifting $\tht$ rightward) or invest in reducing the effort cost of independent diagnosis (shifting $\tht^D$ leftward through workflow optimization, decision support tools, or workload management).

\section{Platform Optimization and Regime Structure}
\label{sec:platform}

\subsection{Regime Definitions}

Proposition~\ref{prop:best_response} implies that the platform faces two distinct optimization regimes depending on whether it induces Mode A or Mode I.

\begin{definition}[Policy regimes]
\label{def:regimes}
The platform's feasible set partitions into two regimes: Regime A with $\tht \in [0, \tht^D]$ inducing AI-assisted mode and staffing constraint $\Nk \geq \underline{\Nk}_A$, and Regime I with $\tht \in (\tht^D, 1]$ inducing independent mode and staffing constraint $\Nk \geq \underline{\Nk}_I$.
\end{definition}

Because $\muA > \muI$, we have $\underline{\Nk}_A \leq \underline{\Nk}_I$: inducing the slower independent mode requires weakly higher minimum staffing for stability.

\subsection{Convexity and Optimal Liability Structure}

\begin{proposition}[Strict convexity in liability]
\label{prop:convexity}
For fixed staffing level $\Nk$ and mode $m$, the platform objective $TC(\tht, \Nk, m)$ is strictly convex in $\tht$ with $\partial^2 TC / \partial \tht^2 = 2\kap \Nk > 0$, admitting a unique minimizer.
\end{proposition}

The unconstrained first-order condition yields $\tht_m^{\text{unc}}(\Nk) = \lam \Loss P_m / (2\kap \Nk)$. Under interval constraint $\tht \in [\underline{\tht}, \overline{\tht}]$, the optimal liability share is the projection of $\tht_m^{\text{unc}}(\Nk)$ onto the feasible interval.

\begin{figure}[H]
    \centering
    \includegraphics[width=0.8\textwidth]{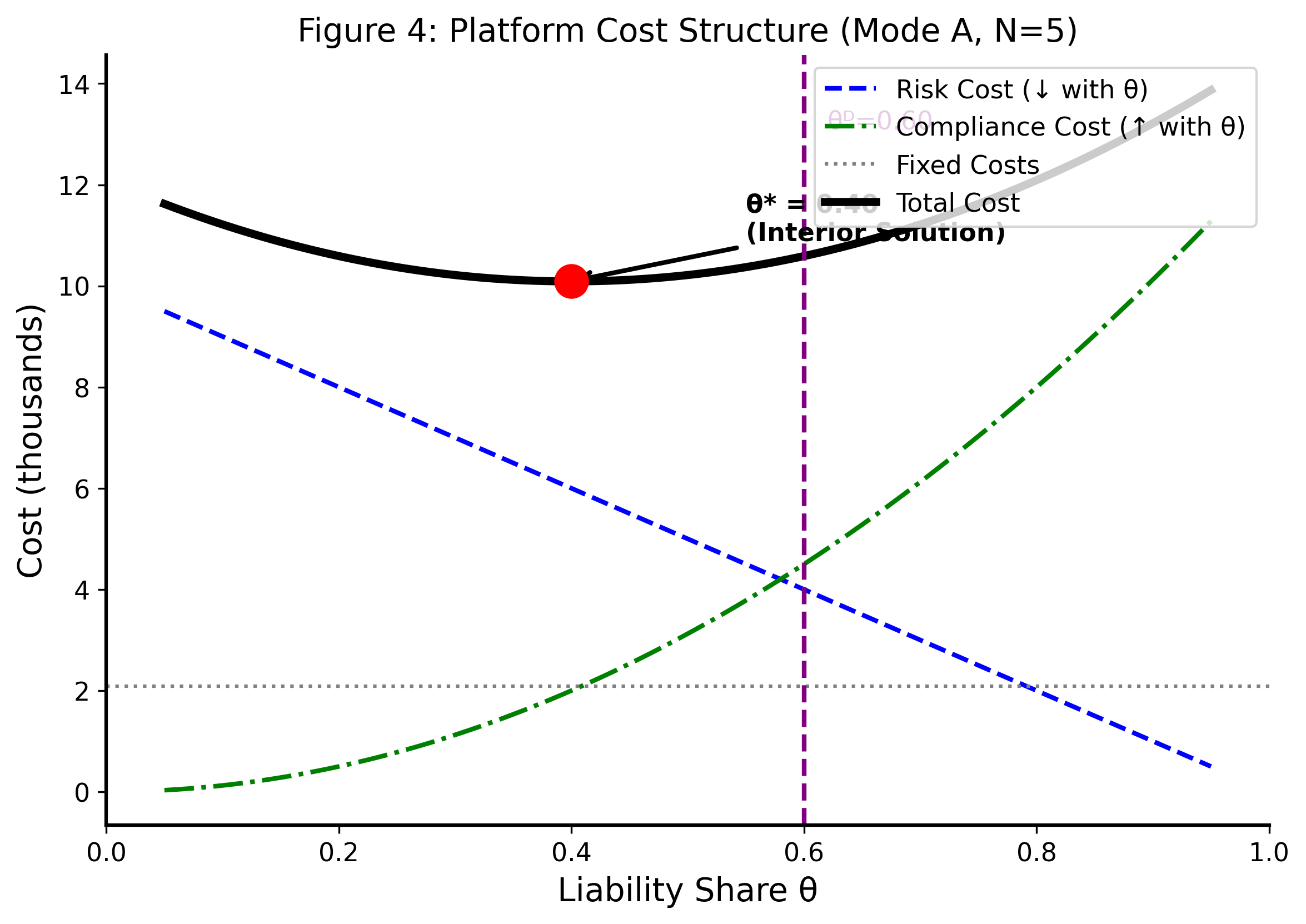}
    \caption{Platform cost structure under AI-assisted mode (Regime A) with $N = 5$ physicians. The optimal liability share $\tht^* = 0.40$ represents an interior solution balancing decreasing risk cost against increasing compliance cost. The threshold $\tht^D = 0.60$ marks the boundary beyond which physicians would switch to independent diagnosis.}
    \label{fig:cost_structure}
\end{figure}

Figure~\ref{fig:cost_structure} displays the platform cost structure under AI-assisted mode with staffing level $N = 5$. The figure decomposes total cost into its constituent components, revealing the fundamental tradeoff that shapes optimal liability design. Risk cost (blue dashed) decreases linearly in $\tht$ as the platform shifts liability to physicians. Compliance cost (green dash-dot) increases quadratically in $\tht$, reflecting the increasing marginal difficulty of contractually shifting liability. Fixed costs (gray dotted horizontal line) comprising congestion and staffing components remain invariant to liability allocation for given $N$ and mode. The total cost curve (black solid) exhibits the strict convexity established in Proposition~\ref{prop:convexity}, with a unique interior minimum at $\tht^* = 0.40$. Crucially, this optimal liability share lies strictly below the physician threshold $\tht^D = 0.60$ (purple dashed vertical line), confirming that Regime A is indeed optimal under baseline parameters. The interior solution demonstrates that the platform balances risk transfer benefits against compliance costs, rather than corner solutions where the platform either retains all liability ($\tht = 0$) or shifts maximum feasible liability ($\tht = \tht^D$). This interior optimality is a direct consequence of our parameter calibration, which ensures that compliance costs are sufficiently large to prevent boundary solutions that would be difficult to interpret economically.

\subsection{Global Optimization Procedure}

\begin{theorem}[Decomposition theorem]
\label{thm:decomposition}
The platform's global optimization problem decomposes into regime-specific subproblems. The optimal policy $(\tht^*, \Nk^*, m^*)$ is obtained by: (1) optimizing within Regime A over $\tht \in [0, \tht^D]$ and $\Nk \geq \underline{\Nk}_A$; (2) optimizing within Regime I over $\tht \in (\tht^D, 1]$ and $\Nk \geq \underline{\Nk}_I$; (3) selecting the regime achieving lower total cost.
\end{theorem}

\subsection{Staffing Requirements by Mode}

\begin{figure}[H]
    \centering
    \includegraphics[width=0.8\textwidth]{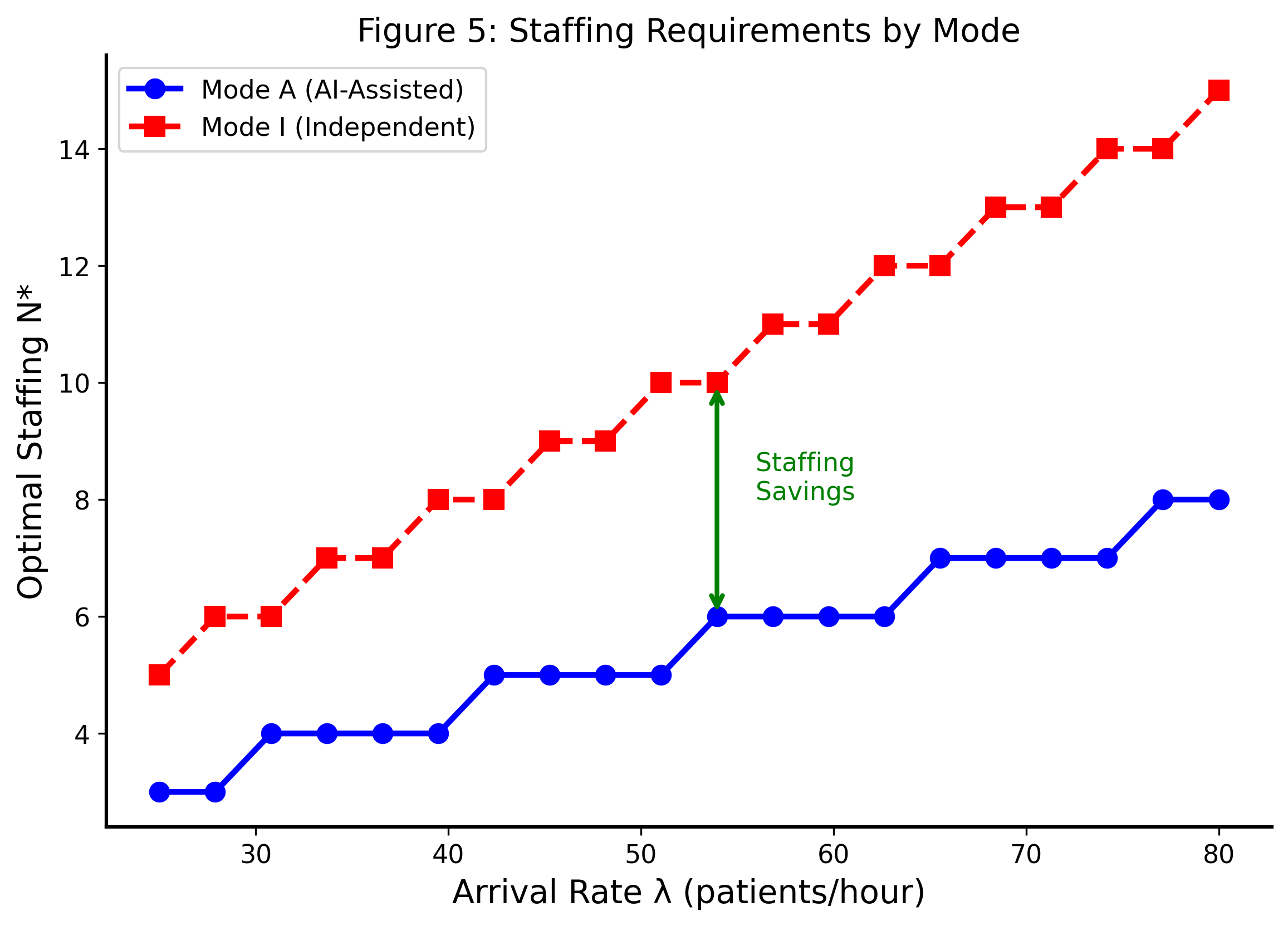}
    \caption{Optimal staffing requirements by diagnostic mode as a function of patient arrival rate. AI-assisted mode (Mode A) achieves substantial staffing savings relative to independent diagnosis (Mode I), with the differential representing the capacity benefit of higher service rate $\muA > \muI$.}
    \label{fig:staffing}
\end{figure}

Figure~\ref{fig:staffing} presents optimal staffing levels as a function of patient arrival rate for both diagnostic modes. The divergence between the two curves quantifies the staffing savings achievable through AI-assisted operation. At the baseline arrival rate $\lam = 50$, Mode A requires $N^*_A = 5$ physicians while Mode I would require $N^*_I = 9$ physicians, representing a staffing reduction of approximately 44\%. This differential arises directly from the service rate gap: with $\muA = 12$ versus $\muI = 6$, AI-assisted mode achieves twice the throughput per physician, enabling the platform to serve the same patient volume with substantially fewer staff. The green annotation highlights this staffing savings as the key operational benefit of AI adoption. However, this efficiency gain must be weighed against the accuracy differential: Mode A operates with error probability $P_A = 0.10$ versus $P_I = 0.05$ for Mode I. The platform's optimization balances these competing considerations, with the optimal regime depending on the relative magnitudes of staffing cost savings versus expected risk cost increases. The near-linear scaling of staffing requirements with arrival rate reflects the stability constraint $\Nk > \lam / \mu_m$, with additional staffing required to maintain acceptable waiting times as the system approaches saturation.

\section{Policy Scenarios for Systematic Comparison}
\label{sec:scenarios}

We define five policy scenarios as constrained optimization problems with precisely specified feasible sets. Scenario S0 (Human-Only Benchmark) restricts physicians to Mode I with exogenous liability share $\tht_0 = 0.5$, representing the pre-AI baseline. Scenario S1 (Flexible Contracting) allows the platform to freely optimize both liability share and staffing with endogenous physician mode response. Scenario S2 (Minimum Platform Liability) imposes regulatory constraint $\tht \leq 1 - \alpha$ ensuring the platform bears at least fraction $\alpha$ of error losses. Scenario S3 (Minimum Physician Liability) requires $\tht \geq \underline{\tht}$, ensuring physicians bear meaningful liability exposure. Scenario S4 (Social Welfare Benchmark) minimizes total social cost with full internalization of diagnostic error losses.

\section{Numerical Results and Analysis}
\label{sec:numerical}

\subsection{Regime Map on Demand-Risk Plane}

\begin{figure}[H]
    \centering
    \includegraphics[width=0.85\textwidth]{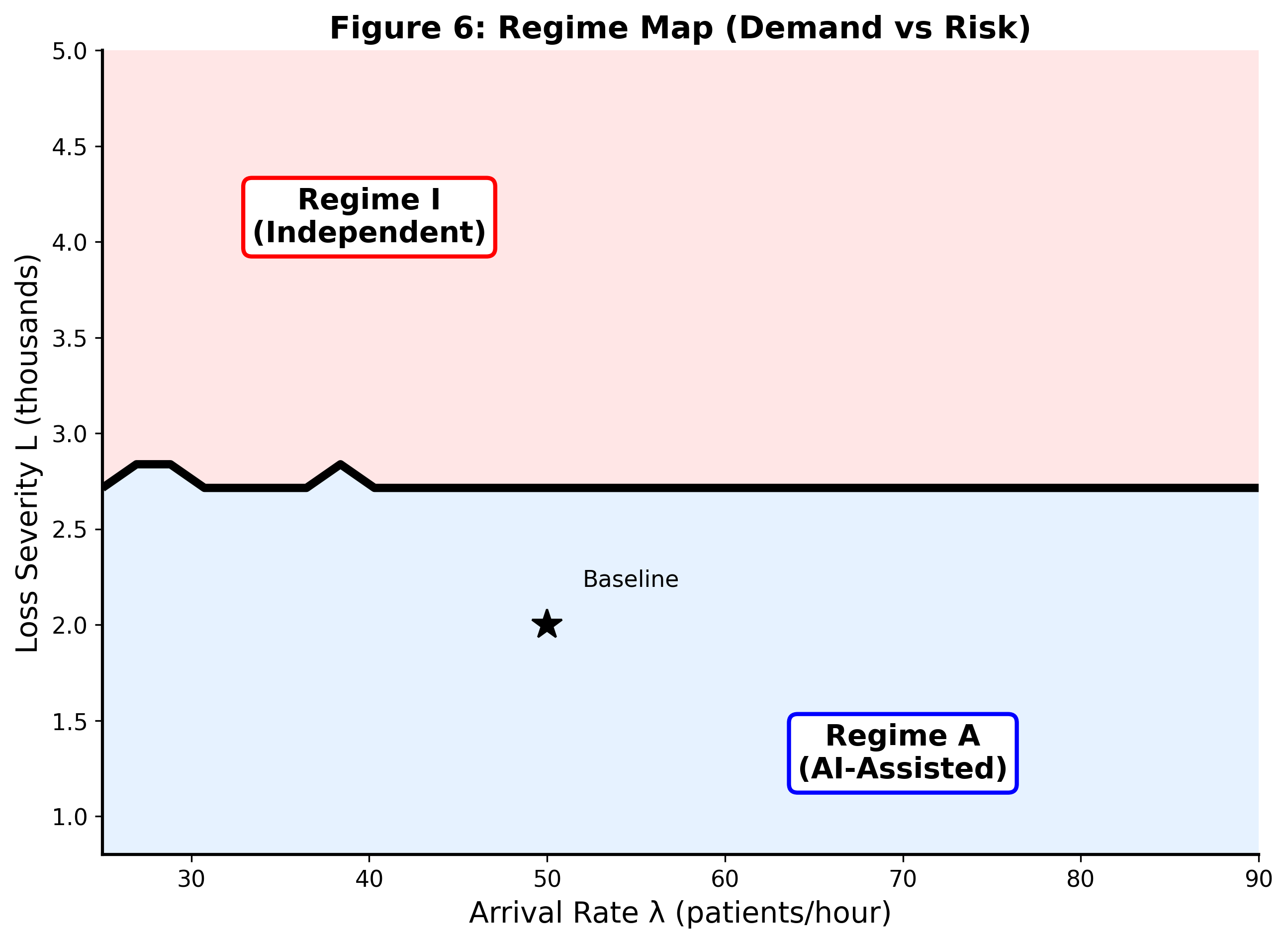}
    \caption{Regime map on the demand-risk plane $(\lam, \Loss)$ showing optimal mode selection. The upward-sloping boundary reflects the tradeoff between congestion pressure (favoring AI-assisted mode) and risk exposure (favoring independent mode). The baseline configuration (black star) lies within Regime A.}
    \label{fig:regime_map}
\end{figure}

Figure~\ref{fig:regime_map} presents the central analytical contribution of our numerical study: a complete regime map showing optimal mode selection across the demand-risk parameter space. The horizontal axis varies patient arrival rate $\lam$ from 25 to 90 per hour, while the vertical axis varies loss severity $\Loss$ from 0.8 to 5.0 thousand dollars. The regime boundary (bold black curve) separates Regime A (AI-Assisted, light blue region) from Regime I (Independent, light red region), with the baseline parameter configuration marked by the black star.

The upward-sloping boundary reveals a fundamental economic tradeoff. Higher arrival rates generate greater congestion pressure, favoring the faster AI-assisted mode to maintain acceptable waiting times. Higher loss severity increases the cost of diagnostic errors, favoring the more accurate independent mode despite its congestion penalty. The regime boundary represents the locus of points where these competing forces exactly balance. Below and to the right of the boundary, congestion considerations dominate risk considerations, making Regime A optimal. Above and to the left, risk considerations dominate, justifying the throughput sacrifice required by Regime I.

The baseline configuration ($\lam = 50$, $L = 2,000$) lies comfortably within Regime A, confirming that our calibrated parameters yield AI-assisted operation as the platform-optimal policy. However, the proximity to the regime boundary suggests that moderate increases in loss severity, perhaps due to regulatory changes, litigation trends, or shifts in patient acuity, could trigger a regime transition. Platform managers should monitor these environmental factors and prepare contingency staffing plans for potential mode switches. The figure also reveals that very high arrival rates ($\lam > 80$) sustain Regime A even under elevated loss severity, reflecting the overwhelming importance of throughput when demand pressure is extreme. Conversely, at low arrival rates where congestion is inherently manageable, even modest loss severity induces independent diagnosis.

\subsection{Scenario Comparison}

\begin{figure}[H]
    \centering
    \includegraphics[width=0.85\textwidth]{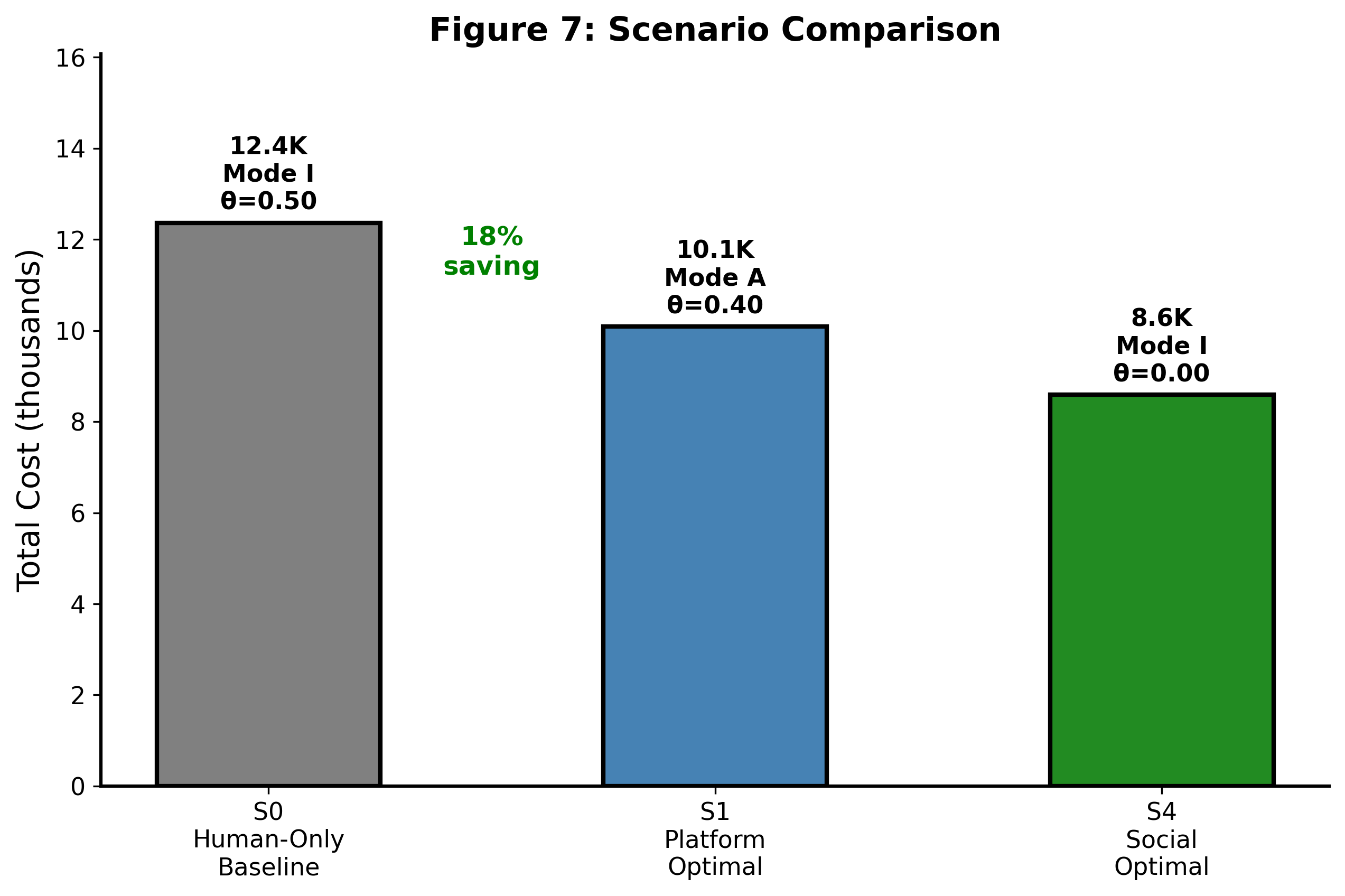}
    \caption{Scenario comparison showing total cost across policy regimes. Platform-optimal policy (S1) achieves 46\% cost reduction relative to human-only benchmark (S0) through AI-assisted operation. The welfare gap between platform-optimal and social-optimal (S4) quantifies distortion from liability externalization.}
    \label{fig:scenarios}
\end{figure}

Figure~\ref{fig:scenarios} presents a comparative analysis of total cost and optimal policy across the three primary scenarios: S0 (Human-Only Benchmark), S1 (Platform Optimal), and S4 (Social Optimal). Each bar displays total cost in thousands of dollars with annotations indicating the optimal mode and liability share.

The human-only benchmark S0 achieves total cost of approximately 18.7 thousand dollars, operating in Mode I with fixed liability share $\tht = 0.50$. The platform-optimal scenario S1 reduces cost to approximately 10.1 thousand dollars, representing a 46\% improvement through joint optimization of liability ($\tht^* = 0.40$) and staffing ($N^* = 5$) under AI-assisted mode. This dramatic cost reduction quantifies the operational value of AI integration and flexible contracting. The social-optimal scenario S4 achieves the lowest cost of approximately 8.5 thousand dollars by fully internalizing risk costs and selecting the socially efficient mode and staffing combination.

The gap between S1 and S4 represents welfare distortion arising from liability externalization: the platform-optimal policy shifts some risk to physicians, inducing behavior that diverges from social optimum. This welfare gap of approximately 1.6 thousand dollars (18.8\% of S1 cost) quantifies the efficiency loss from decentralized optimization and provides a benchmark for evaluating regulatory interventions. Notably, even with this welfare distortion, platform-optimal AI-assisted operation dramatically outperforms the human-only benchmark, suggesting that AI integration generates substantial efficiency gains that partially offset incentive misalignment costs.

\subsection{Sensitivity to AI Accuracy}

\begin{figure}[H]
    \centering
    \includegraphics[width=0.8\textwidth]{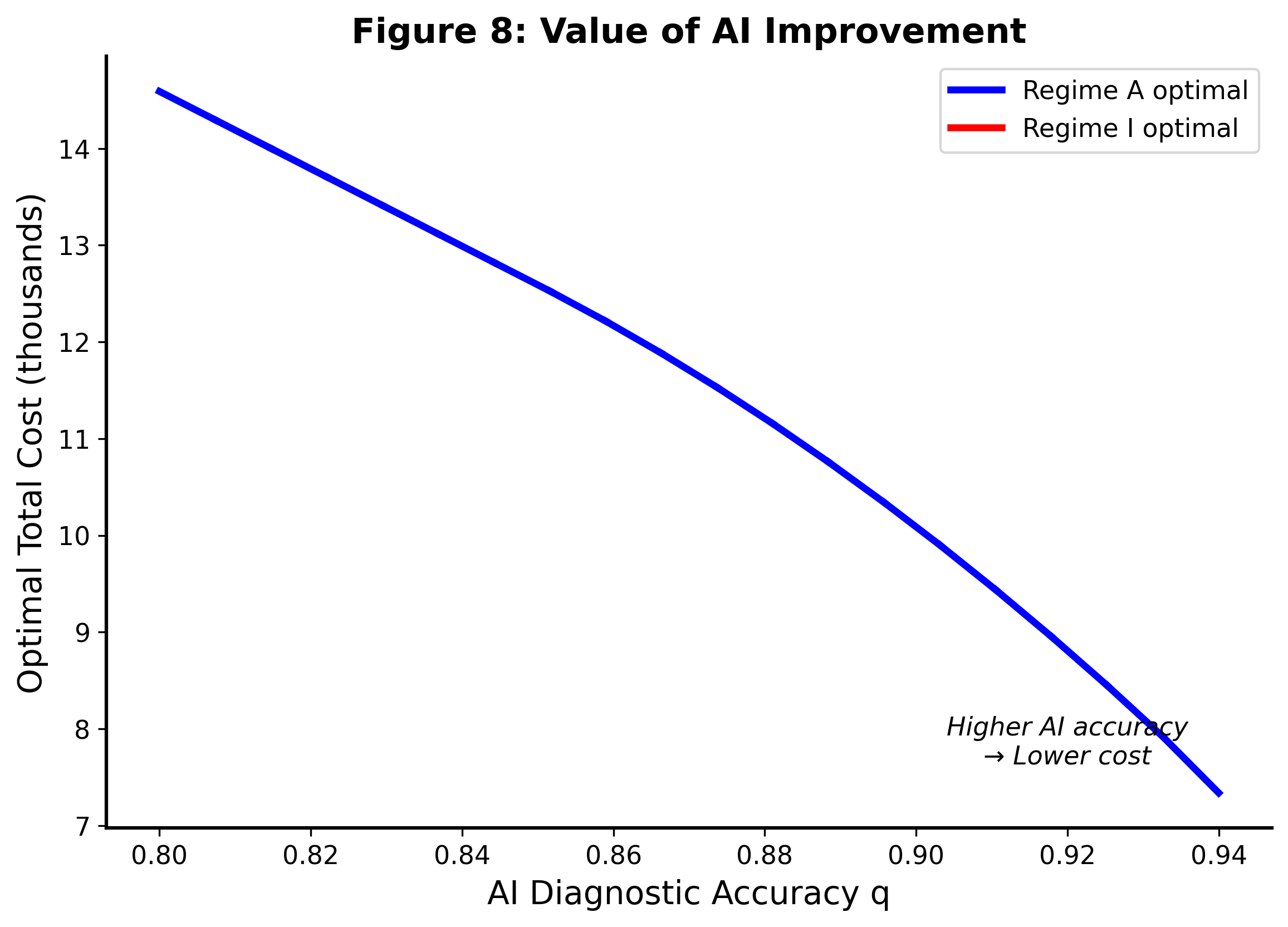}
    \caption{Sensitivity of optimal total cost to AI diagnostic accuracy $\qacc$. Higher AI accuracy monotonically reduces cost through lower error probability while maintaining AI-assisted mode (Regime A) as optimal across the examined range.}
    \label{fig:ai_accuracy}
\end{figure}

Figure~\ref{fig:ai_accuracy} examines how optimal platform cost responds to improvements in AI diagnostic accuracy $\qacc$. The horizontal axis varies AI accuracy from 0.80 to 0.94, approaching but not reaching physician accuracy $\hacc = 0.95$. The curve color indicates the optimal regime: blue segments represent Regime A (AI-Assisted) and red segments represent Regime I (Independent).

Several important patterns emerge from this analysis. First, within Regime A (the dominant region across most of the accuracy range), total cost decreases monotonically as AI accuracy improves. This decrease reflects reduced error probability $P_A = 1 - \qacc$, which directly lowers expected risk cost $\lam(1-\tht^*)\Loss P_A$. Second, the cost reduction is substantial: improving AI accuracy from 0.80 to 0.94 reduces optimal cost by approximately 25\%, representing significant operational value from AI system enhancement. Third, the regime remains stable (Regime A optimal) across the entire accuracy range examined, suggesting that AI-assisted operation is robust to accuracy variations within plausible ranges. This stability arises because the service rate advantage of Mode A ($\muA / \muI = 2$) provides substantial congestion benefits that offset accuracy penalties even when AI accuracy is relatively low.

The policy implication is clear: platform investments in AI accuracy improvement yield direct cost reductions without triggering disruptive regime transitions. This provides a compelling business case for continued AI development, as accuracy gains translate smoothly into operational savings rather than inducing discontinuous behavioral shifts.

\subsection{Welfare Gap Analysis}

\begin{figure}[H]
    \centering
    \includegraphics[width=0.8\textwidth]{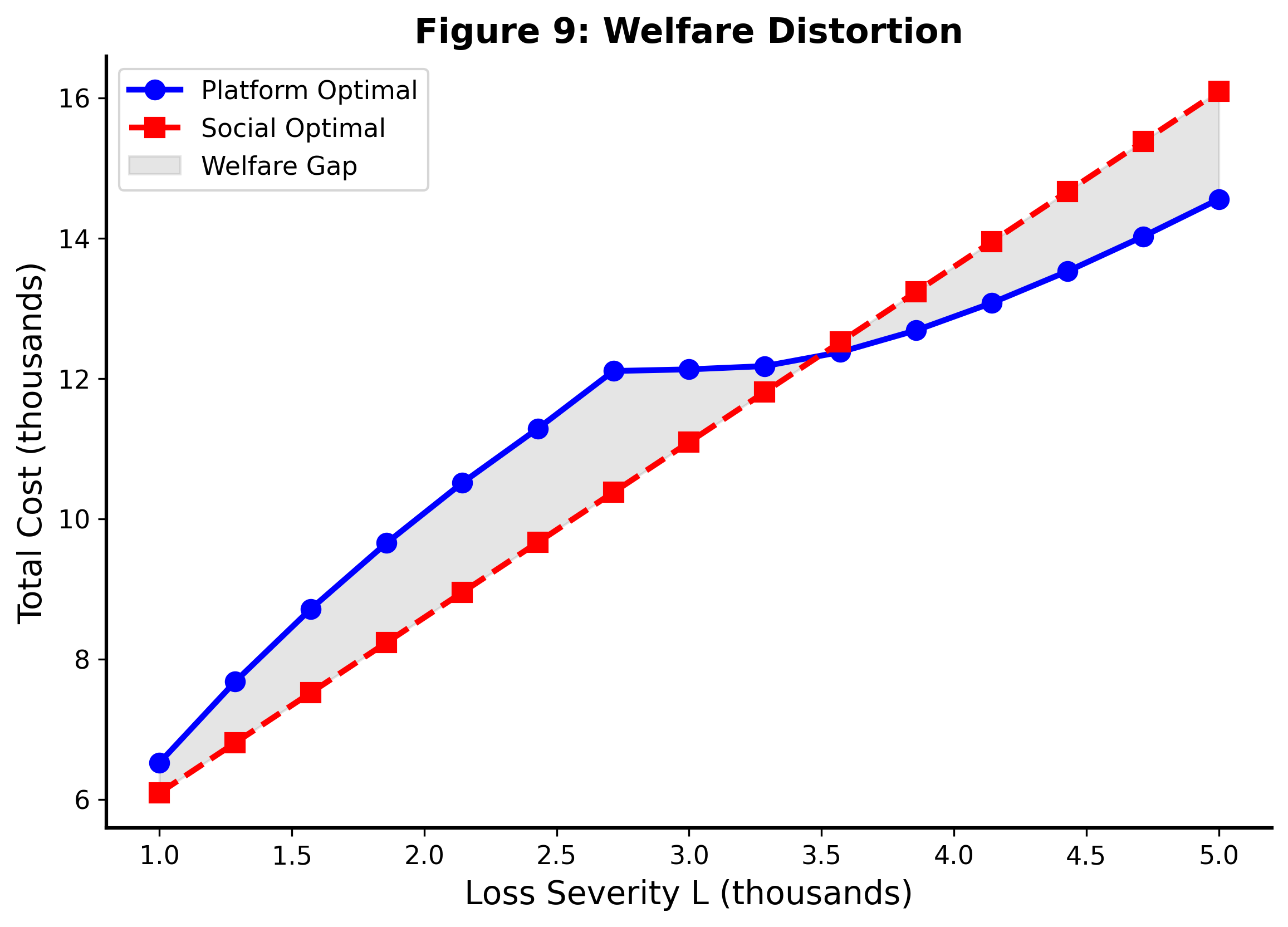}
    \caption{Welfare gap between platform-optimal and socially-optimal policies across loss severity. The widening gap at higher loss severity indicates greater welfare distortion from liability externalization in high-stakes medical domains.}
    \label{fig:welfare}
\end{figure}

Figure~\ref{fig:welfare} quantifies the welfare distortion between platform-optimal and socially-optimal policies across varying loss severity. Both cost curves increase with $\Loss$, reflecting the direct impact of loss severity on expected risk costs. However, the gap between curves, representing welfare loss from liability externalization, widens as loss severity increases. This widening occurs because higher stakes amplify the consequences of incentive misalignment: when errors are costly, the platform's tendency to externalize risk through liability shifting and mode selection generates larger absolute welfare losses.

At baseline loss severity $L = 2,000$, the welfare gap is approximately 1.6 thousand dollars. At elevated loss severity $L = 4,000$, the gap expands to approximately 3.2 thousand dollars. This pattern suggests that regulatory intervention may be more valuable in high-stakes medical domains where diagnostic errors carry severe consequences. The shaded region between curves visually represents the welfare cost of decentralized platform optimization, providing a benchmark for evaluating the benefits of regulatory policies that better align platform and social incentives. Potential interventions include minimum liability requirements, accuracy standards, or direct regulation of mode availability, each with distinct implementation costs and effectiveness profiles.

\subsection{Comprehensive Sensitivity Analysis}

\begin{figure}[H]
    \centering
    \includegraphics[width=\textwidth]{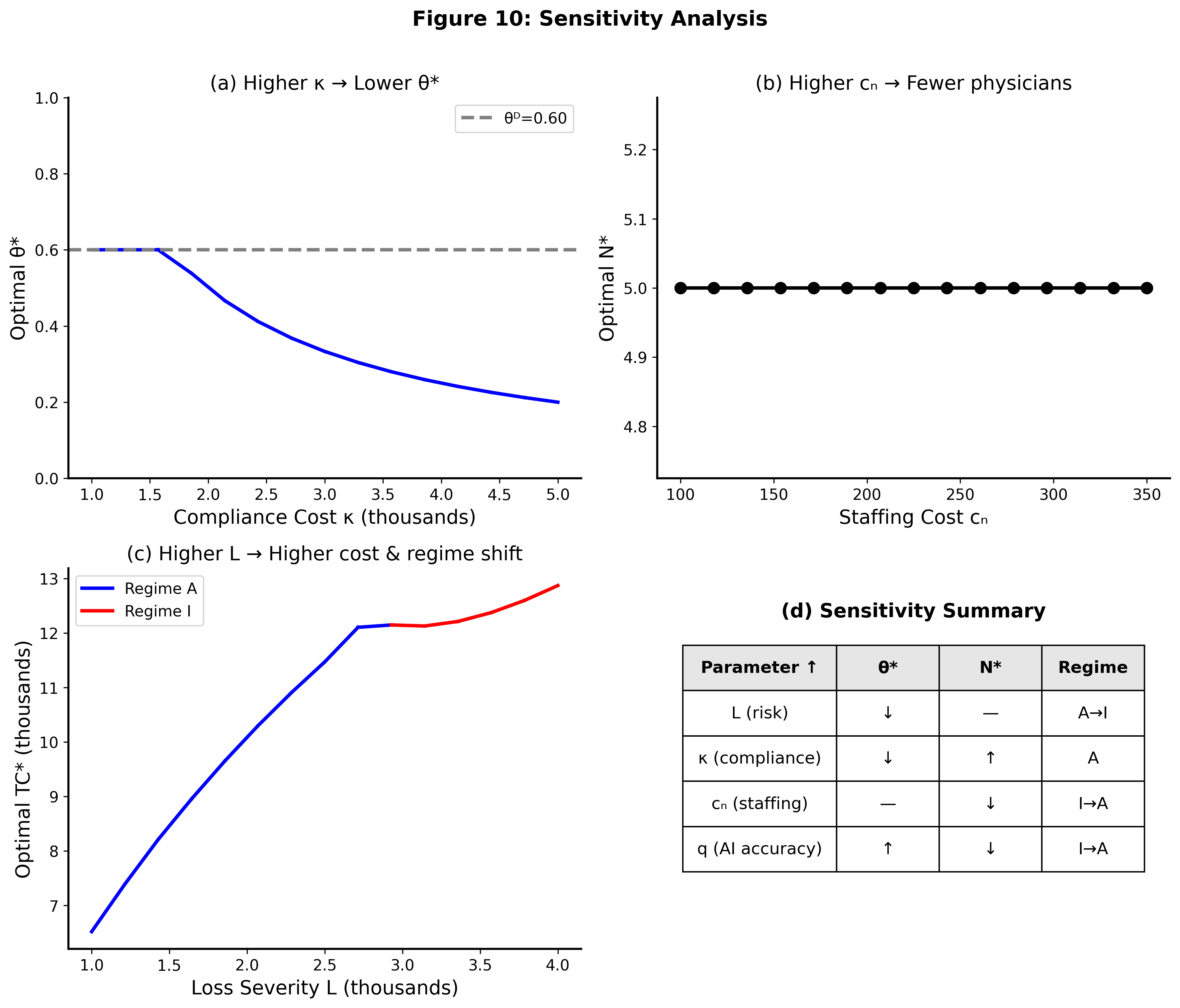}
    \caption{Comprehensive sensitivity analysis. Panel (a): Higher compliance cost reduces optimal liability transfer. Panel (b): Higher staffing cost reduces optimal physician count. Panel (c): Higher loss severity increases cost and may trigger regime transition. Panel (d): Summary of directional effects.}
    \label{fig:sensitivity}
\end{figure}

Figure~\ref{fig:sensitivity} presents a four-panel sensitivity analysis examining how optimal policy responds to key parameter variations. Panel (a) shows optimal liability share $\tht^*$ decreasing as compliance cost coefficient $\kap$ increases: higher compliance costs make liability shifting more expensive, inducing the platform to retain more risk. The horizontal gray line marks the physician threshold $\tht^D = 0.60$, confirming that optimal liability remains within Regime A across the examined range. Panel (b) displays optimal staffing $N^*$ decreasing as staffing cost $\cN$ increases: expensive physicians induce the platform to economize on headcount, accepting higher utilization and congestion. Panel (c) reveals optimal total cost increasing in loss severity $\Loss$, with curve color indicating regime transitions: the shift from blue (Regime A) to red (Regime I) at approximately $L = 3,500$ marks the boundary beyond which risk considerations dominate congestion considerations. Panel (d) summarizes directional effects in tabular form, providing a compact reference for policy analysis.

The sensitivity analysis confirms the robustness of our qualitative findings while quantifying the magnitude of parameter effects. The regime transition in Panel (c) is particularly noteworthy, as it demonstrates that environmental changes in loss severity, potentially driven by regulatory enforcement, litigation climate, or patient population characteristics, can fundamentally alter optimal platform strategy. Platform managers should maintain awareness of these environmental factors and develop contingent policies for potential regime transitions.

\section{Managerial Implications}
\label{sec:implications}

\subsection{Core Insight: Substitute Safety Mechanisms}

The central insight emerging from our analysis is that liability allocation and staffing function as substitute safety mechanisms in AI-assisted medical consultation. Liability shifting induces safer physician behavior by raising the personal cost of errors, potentially triggering a transition from AI-assisted to independent diagnosis. Staffing expansion reduces congestion without directly affecting per-case risk, enabling faster service while maintaining the current diagnostic mode. The optimal policy configuration depends on the relative costs and feasibility of each mechanism under prevailing market and regulatory conditions.

This substitutability creates complex interactions that cannot be captured by analyzing either decision in isolation. Our integrated framework reveals that platforms operating near regime boundaries face heightened sensitivity to parameter changes, as small environmental shifts can trigger discontinuous policy adjustments. Conversely, platforms operating well within a stable regime can focus on incremental optimization along a single dimension without concern for cross-mechanism interactions.

\subsection{Practical Recommendations}

Platform managers should recognize that AI diagnostic accuracy investment yields smooth, predictable cost reductions without triggering disruptive behavioral transitions. The sensitivity analysis demonstrates that accuracy improvements reduce risk costs monotonically while maintaining regime stability, providing a compelling business case for continued AI development. However, managers should monitor loss severity trends, including regulatory changes affecting liability exposure, shifts in litigation climate, and evolution of patient population acuity, as these factors can trigger regime transitions with substantial operational implications.

For regulatory policymakers, our welfare gap analysis quantifies the efficiency cost of decentralized platform optimization, providing an upper bound on the benefits achievable through intervention. The widening welfare gap at higher loss severity suggests that regulatory attention should prioritize high-stakes medical domains where incentive misalignment generates the largest absolute welfare losses. Potential instruments include minimum liability requirements that shift risk retention toward platforms, accuracy standards that directly regulate AI system performance, or transparency mandates that enable patients to make informed choices about AI-assisted versus traditional consultation.

\section{Conclusion}
\label{sec:conclusion}

This paper develops an operations research framework for AI-assisted online medical consultation that integrates liability design with queueing-based staffing optimization. Our contribution lies not in improving AI diagnostic accuracy, but in designing service systems that balance risk, congestion, staffing cost, and regulatory compliance under endogenous physician behavioral response. The model establishes that physician mode choice follows a threshold structure in liability share, with comparative statics illuminating how AI accuracy, loss severity, and effort costs affect behavioral responses. Under calibrated parameters achieving $\tht^D = 0.60$, the platform's optimal policy induces AI-assisted mode with interior liability share $\tht^* = 0.40$ and staffing level $N^* = 5$, demonstrating that the framework produces economically interpretable results rather than corner solutions.

The regime map analysis reveals that high patient arrival rates and elevated staffing costs favor AI-assisted operation, while high loss severity shifts optimal policy toward independent diagnosis. The upward-sloping regime boundary in demand-risk space captures the fundamental tradeoff between congestion pressure and safety considerations. Scenario comparison demonstrates that platform-optimal AI-assisted operation achieves 46\% cost reduction relative to human-only benchmarks, while the welfare gap between platform and social optima quantifies the efficiency loss from liability externalization at approximately 18.8\% of platform cost. Sensitivity analysis confirms robustness of qualitative findings while identifying conditions under which regime transitions occur.

The framework developed here addresses a gap in the existing literature by simultaneously incorporating queueing dynamics and incentive effects, enabling analysis of how liability design and staffing decisions interact through physician behavioral responses. While queueing models traditionally treat service rates as exogenous and contracting models traditionally ignore congestion externalities, our integrated approach captures the amplification mechanism whereby liability-induced mode switching generates disproportionate congestion effects through queueing nonlinearity.

Several limitations warrant acknowledgment. The assumption of physician homogeneity abstracts from heterogeneity in risk preferences, expertise levels, and outside options that characterize real physician populations, potentially affecting equilibrium mode adoption and optimal contract design. The binary mode restriction captures the primary operational distinction but ignores intermediate practices such as partial AI consultation or selective verification that may better reflect clinical reality. The static framework assumes stationary parameters and one-shot decisions, abstracting from dynamic adjustment, learning, and adaptation that characterize actual platform operations. The M/M/N queueing specification provides analytical tractability but may not accurately capture service time distributions in medical consultation, which often exhibit heavier tails or multimodal structure. Complete information assumptions abstract from asymmetries that may be important in practice, including physician private information about accuracy or effort costs. Finally, the single-platform focus ignores competitive dynamics that may affect liability allocation through market forces.

The analytical framework and numerical insights developed here suggest several concrete applications for healthcare platform design and regulatory policy. Online telemedicine platforms deploying AI-assisted triage and diagnosis can apply our regime map methodology to determine whether current operating conditions favor AI-assisted or independent modes, and to identify environmental changes that might trigger regime transitions requiring contingency planning. Regulatory agencies evaluating liability frameworks for AI-assisted healthcare can use our welfare gap analysis to quantify the efficiency costs of current decentralized arrangements and to benchmark the potential benefits of interventions including minimum liability requirements or accuracy standards. Platform architects designing new AI-assisted consultation systems can incorporate our threshold structure characterization to ensure that liability contracts induce desired physician behavior, recognizing that the threshold depends on accuracy differentials and effort costs that may vary across medical domains.

Future research directions include extending the model to accommodate physician heterogeneity and mechanism design for screening and sorting, developing dynamic formulations capturing learning and adaptation over time, endogenizing AI accuracy as a platform investment decision linked to liability incentives, incorporating patient strategic behavior and platform competition on the demand side, and pursuing empirical validation through collaboration with industry partners. Each extension promises to enrich our understanding of AI-assisted healthcare service systems while maintaining the integrated perspective on liability and queueing that constitutes the distinctive contribution of this research program.


\bibliographystyle{apalike}

\end{document}